\documentclass[12pt]{article}
\usepackage{epsfig}
\begin {document}
\parindent=15pt
\begin{center}
{\bf THE ROLE OF INCIDENT PARTON TRANSVERSE MOMENTA IN HEAVY QUARK
HADROPRODUCTION}\\
\vspace{.4cm}
Yu.M.Shabelski \\
\vspace{.4cm}
Petersburg Nuclear Physics Institute, \\
Gatchina, St.Petersburg 188350 Russia \\
\end{center}

\vspace{.1cm}
{\it Talk, given at HERA Monte Carlo Workshop, \\ 27-30 April 1998,
DESY, Hamburg}

\vspace{.1cm}

\begin{abstract}
The conventional NLO parton model is enough for the description of total
cross sections and one-particle distributions. In the case of
two-particle correlations, the collinear approximation has failed, and
it is necessary to account for the transverse momenta of initial
partons. The different possibilities to do this are discussed.
\end{abstract}

\vspace{.1cm}

E-mail: SHABEL@MAIL.DESY.DE \\

\section{Introduction}
The investigation of heavy quarks production in high energy hadron
collisions provides a method for studying the internal structure of
hadrons. Some problems are the same in hadroproduction and
photo-/electroproduction processes. So, the review of the situation of
the heavy quark hadroproduction can be useful for the interpretation of
HERA data.

In this talk we present a short review of heavy quark hadroproduction.
The theoretical predictions are usually obtained in the NLO parton model
\cite{1}. The assumptions which are used for simplifications of the
computations are considered in Sect. 2. In the case of one-particle
distributions even LO ($\sim \alpha_s^2$) parton model with collinear
approximation is enough for the data description, NLO contributions
($\sim \alpha_s^3$) only change the normalizations. On the other hand,
in the case of two-particle distributions, see Sect. 3, the collinear
approximation has failed, and it is necessary to account for the
transverse momenta of the incident partons. The possibility to include
the transverse momenta of the incident partons in the framework of
semihard theory \cite{GLR}, where the virtualities and polarizations of
the gluons are taken into account, is considered in Sect. 4. In Ref.
\cite{3} we presented the results for main and simplest subprocess,
$gg \rightarrow \overline{Q} Q \; (\sim \alpha_{s}^{2})$ for
hadroproduction, and
$\gamma g \rightarrow \overline{Q} Q \; (\sim \alpha_{s})$ for photo-
and electroproduction.

\section{Conventional NLO parton model}

The conventional NLO parton model expression for the heavy quark
hadroproduction cross sections has the factorization form \cite{CSS}:
\begin{equation}
\sigma (a b \rightarrow Q\overline{Q}) =
\sum_{ij} \int dx_i dx_j G_{a/i}(x_i,\mu_F) G_{b/j}(x_j,\mu_F)
\hat{\sigma} (i j \rightarrow Q \overline{Q}) \;,
\label{pm}
\end{equation}
where $G_{a/i}(x_i,\mu_F)$ and $G_{b/j}(x_j,\mu_F)$ are the structure
functions of partons $i$ and $j$ in the colliding hadrons $a$ and $b$,
$\mu_F$ is the factorization scale (i.e. virtualities of incident
partons) and $\hat{\sigma} (i j \rightarrow Q \overline{Q})$ is
the cross section of the subprocess which is calculated in
perturbative QCD. The last cross section can be written as a sum of
LO and NLO contributions, $\hat{\sigma} (i j \rightarrow Q\overline{Q})
= \alpha_s^2(\mu_R) \sigma^{(o)}_{ij} + \alpha_s^3(\mu_R)
\sigma^{(1)}_{ij}$, where $\mu_R$ is the renormalization scale, and
$\sigma^{(o)}_{ij}$ as well as $\sigma^{(1)}_{ij}$ depend practically
only on one variable $\rho = \frac{4m_Q^2}{\hat{s}}$ ,
$\hat{s} = x_i x_j s_{ab}$.

The expression (1) corresponds to the process shown schematically in
Fig. 1 with
\begin{equation}
q_{1T} = q_{2T} =0 \;.
\end{equation}
The main contribution to the cross section at small $x$ is
known to come from gluon-gluon fusion, $i = j = g$.

\begin{figure}[htb]
\begin{center}
\mbox{\psfig{file=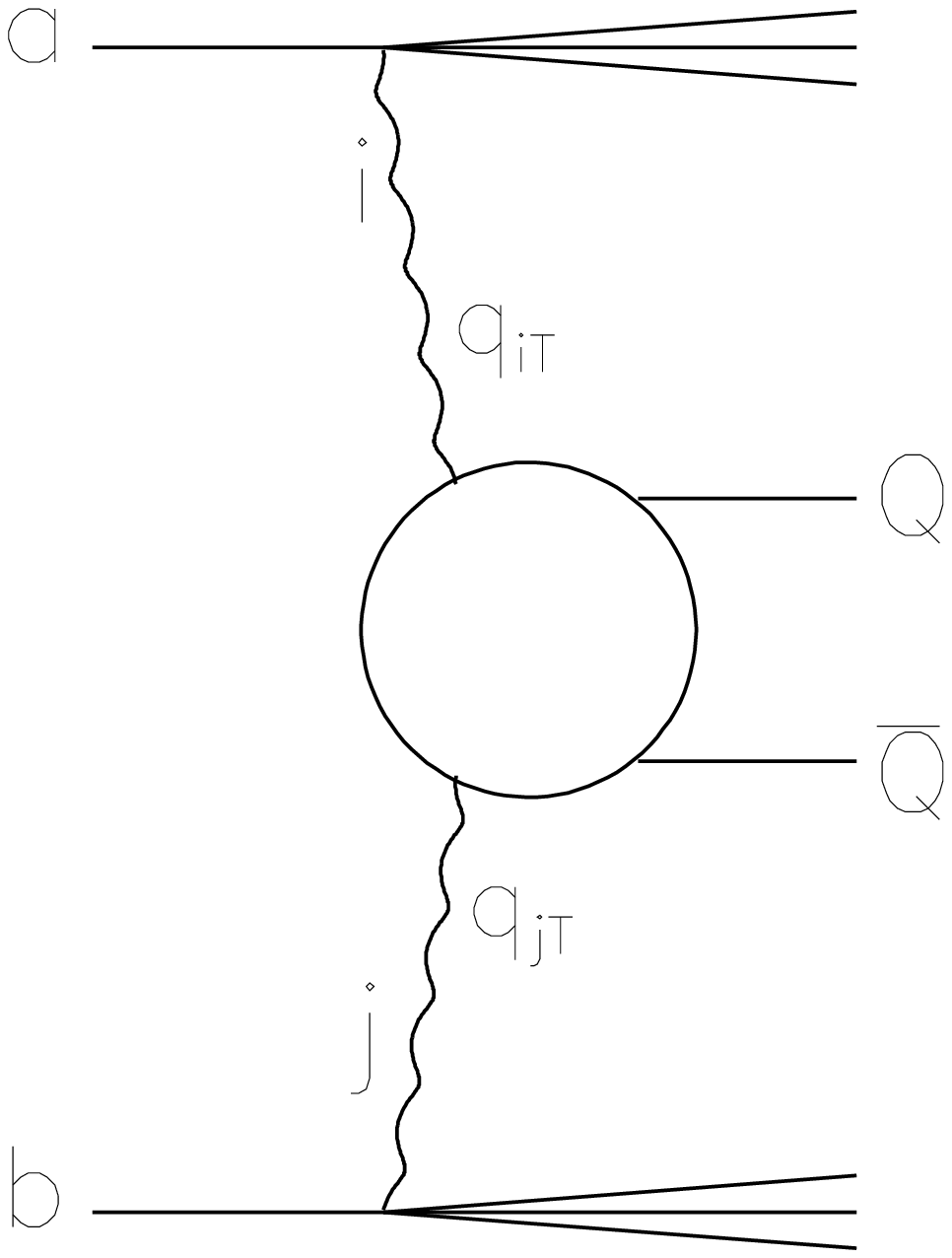,width=0.35\textwidth}} \\
Fig. 1. Heavy quark production in parton model.
\end{center}
\end{figure}

The principal uncertainties of any numerical QCD calculation of heavy
flavour production are connected with the unknown values of the
parameters: both scales, $\mu_F$ and $\mu_R$\footnote{These
uncertainties should disappear when one sums up all the high order
contributions. Sometimes people say that strong scale dependence of the
calculated results in LO or NLO means the large contribution of high
order diagrams and weak dependence means their small contribution. Of
course, it is not true. Strong scale dependence of NLO results means
only strong scale dependence of high order contributions but at some
fixed scale value the last ones can be numerically small. Weak scale
dependence of NLO results means weak scale dependence of high order
terms but they can be numerically large.}, and the exact value of heavy
quark mass, $m_Q$. The values of both scales should be of the order of
hardness of the considered process, however nobody can say what is
better to use for scales, $m_Q$, $m_T = \sqrt{m_Q^2 + p_T^2}$ or
$\hat{s}$. The phenomenological parton densities are sometimes (at very
small $x$) in contradiction \cite{ASS} with the general properties of
perturbative QCD. However it is just the region that dominates in the
heavy quark production at high energies\footnote{In the case of charm
production, $m_c$ = 1.4GeV, at LHC, $\sqrt{s}$ = 14 TeV, the product
$x_1x_2$ of two gluons (both $x_1$ and $x_2$ are the integral variable)
is equal to $4\cdot 10^{-8}$.}. Another problem of parton model is the
collinear approximation. The transverse momenta of the incident partons,
$q_{iT}$ and $q_{jT}$ are assumed to be zero, and their virtualities are
accounted for only via structure functions; the cross sections
$\sigma^{(o)}_{ij}$ and $\sigma^{(1)}_{ij}$ are assumed to be
independent on these virtualities.

The NLO parton model calculations of the total cross sections of
$c\bar{c}$ and $b\bar{b}$ production, as functions of the beam
energy, for $\pi^- N$ and $p-N$ collisions can be found in \cite{FMNR}.
These results depend strongly (on the level of several times) on the
numerical values of quark masses as well as on the both scales, $\mu_F$
and $\mu_R$. Some experimental data are in contradiction with each
other, however generally they are in agreement with NLO parton model
predictions.

\begin{figure}[htb]
\begin{center}
\mbox{\psfig{file=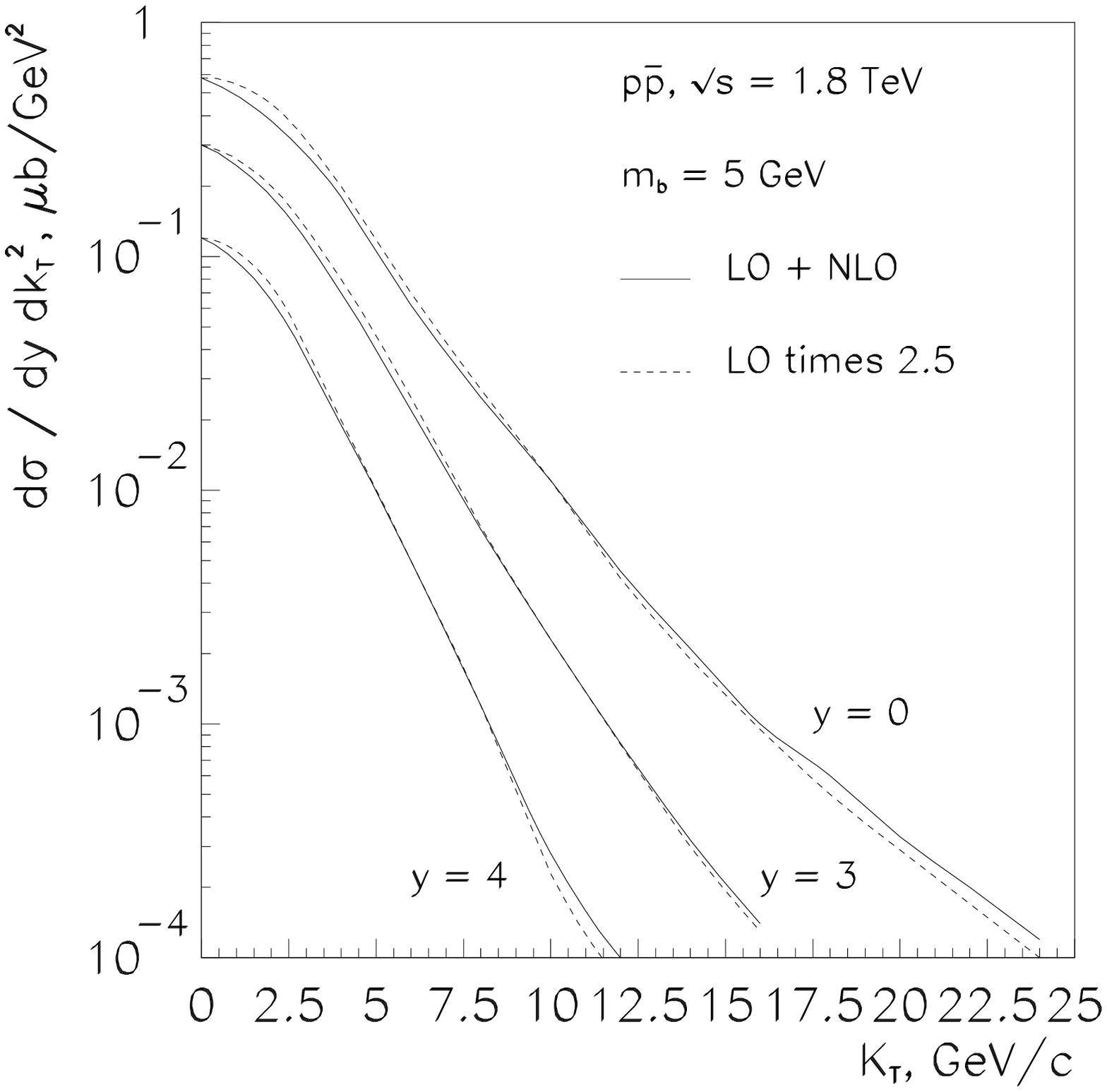,width=0.50\textwidth}} \\
Fig. 2. $p_T$-distributions for $p\bar{p} \rightarrow b + X$ at
$\sqrt{s}$ = 1.8 TeV, for different values of rapidity.
\end{center}
\end{figure}

The NLO contributions to one-particle distributions lead only to
renormalization of LO results, practically without correction of the
shapes of a distributions \cite{NDE1,MNR}. It means that instead of
more complicate calculation of $p_T$, or rapidity distributions, in NLO,
it is enough to calculate them in LO, and multiply after by K-factor
\begin{equation}
K = \frac{LO + NLO}{LO} \;,
\end{equation}
which can be taken, say, from the results for total production cross
sections. The comparison of LO + NLO calculations with LO multiplied by
K-factor is presented in Fig. 2 taken from Ref. \cite{NDE1}.

The values of K-factors and their energy and scale dependences for
several sets of structure functions were calculated in Refs.
\cite{SCG,LSCSh}.

The experimental data for $x_F$-distributions of D-mesons produced in 
$\pi N$ interactions \cite{Adam,Alv} are in agreement with the parton 
model distributions for bare quarks, as one can see in Fig. 3 taken from 
\cite{FMNR}. It means that the fragmentation processes are not important 
here, or they are compensated by, say, recombination processes. The 
shape of $x_F$-distributions does not depend practically on the mass of 
$c$-quark.

\begin{figure}
\begin{center}
\centerline{\epsfig{figure=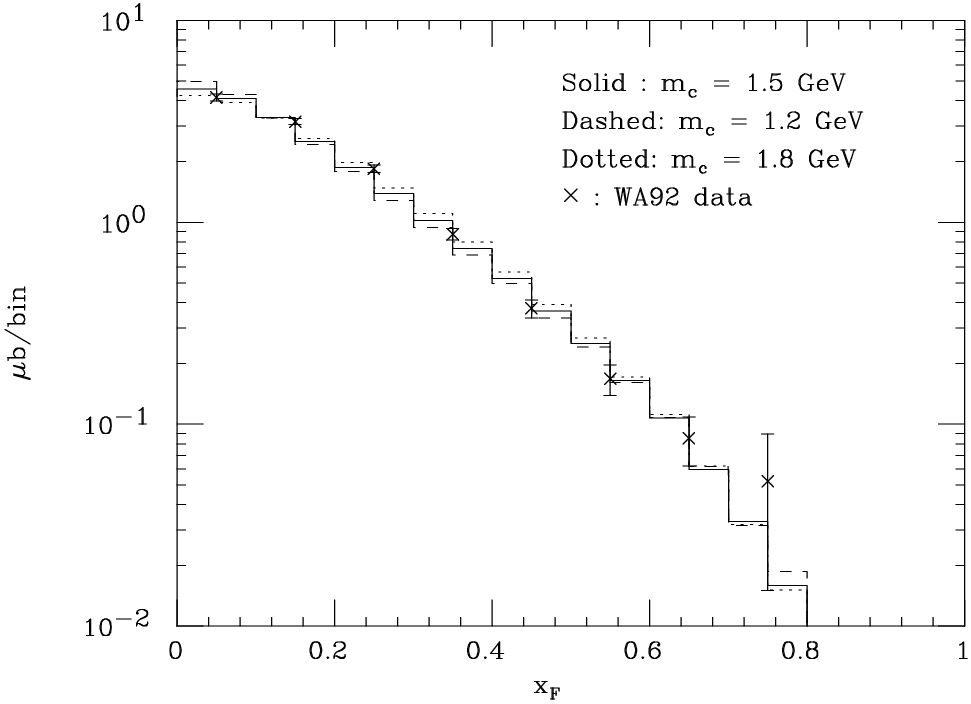,width=0.45\textwidth,clip=}
            \hspace{0.3cm}
            \epsfig{figure=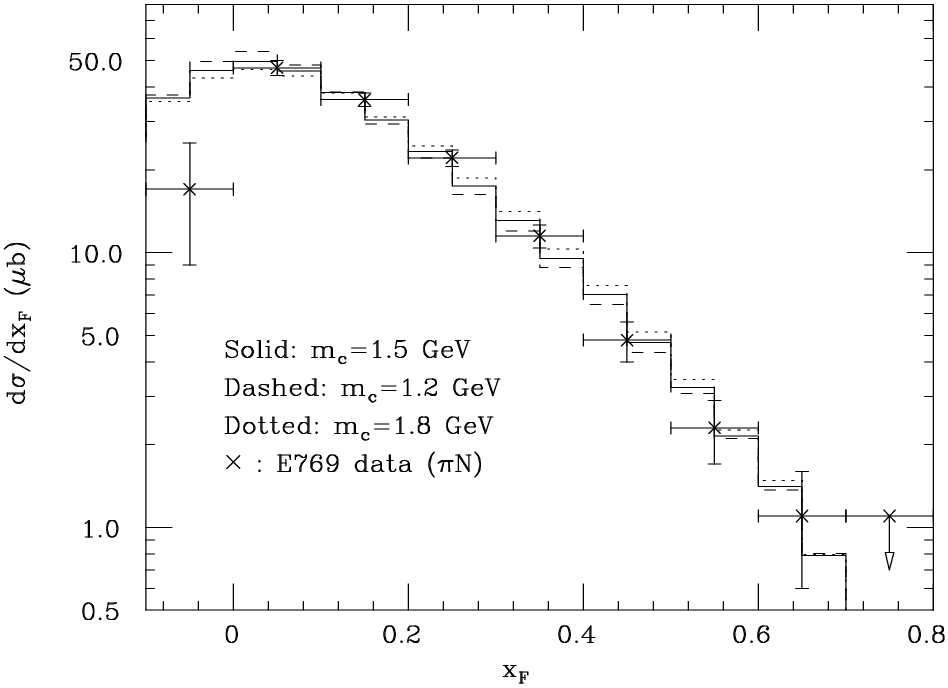,width=0.45\textwidth,clip=}}
Fig. 3. Experimental $x_F$ distributions for $D$ mesons, compared to the
NLO parton model prediction for charm quarks.
\end{center}
\end{figure}

The data on one-particle $p_T$-distributions,including the hadronic
colliders data for the case of beauty production, also can be described 
by the NLO parton model, see \cite{FMNR}.

\section{Azimuthal correlations and failure of the collinear
approximation}

The azimuthal angle $\phi$ is defined as an opening angle between two
produced heavy quarks, projected onto a plane perpendicular to the beam.
In the LO parton model this angle between them is exactly $180^o$. In
the case of NLO parton model a distribution over $\phi$ angle appears
\cite{MNR}.

The investigation of such distributions is very important. In
one-partricle distributions, the sum of LO and NLO contributions of the
parton model practically coinsides with the LO contribution multiplied
by $K$-factor. So we can not control the magnitudes of LO and NLO
contributions separately. In the case of azimuthal correlations all
difference from the trivial $\delta(\phi - \pi)$ distribution comes from
NLO contribution.

The experimental data on azimuthal correlations are claimed (see
\cite{BEAT} and Refs. therein) to be in disagreement with the NLO
predictions, for the cases of charm pair hadro- and photoproduction at
fixed target energies. The level of disagreements can be seen in Fig. 4
(solid histograms) taken from Ref. \cite{FMNR}. These data can be
described \cite{FMNR}, assuming the comparatively large intrinsic
transverse momenta of incoming partons ($k_T$ kick). For each event, in
the longitudinal centre-of-mass frame of the heavy quark pair, the
$Q\overline{Q}$ system is boosted to rest. Then a second transverse
boost is performed, which gives the pair a transverse momentum equal to
$\vec{p}_T(Q\overline{Q}) + \vec{k}_T(1) + \vec{k}_T(2)$;
$\vec{k}_T(1)$ and $\vec{k}_T(2)$ are the transverse momenta of the
incoming partons, which are chosen randomly, with their moduli
distributed according to
\begin{equation}
\frac{1}{N}\frac{dN}{dk_T^2}=\frac{1}{\langle k_T^2 \rangle}
      \exp(-k_T^2/\langle k_T^2 \rangle).
\end{equation}

\begin{figure}
\begin{center}
\centerline{\epsfig{figure=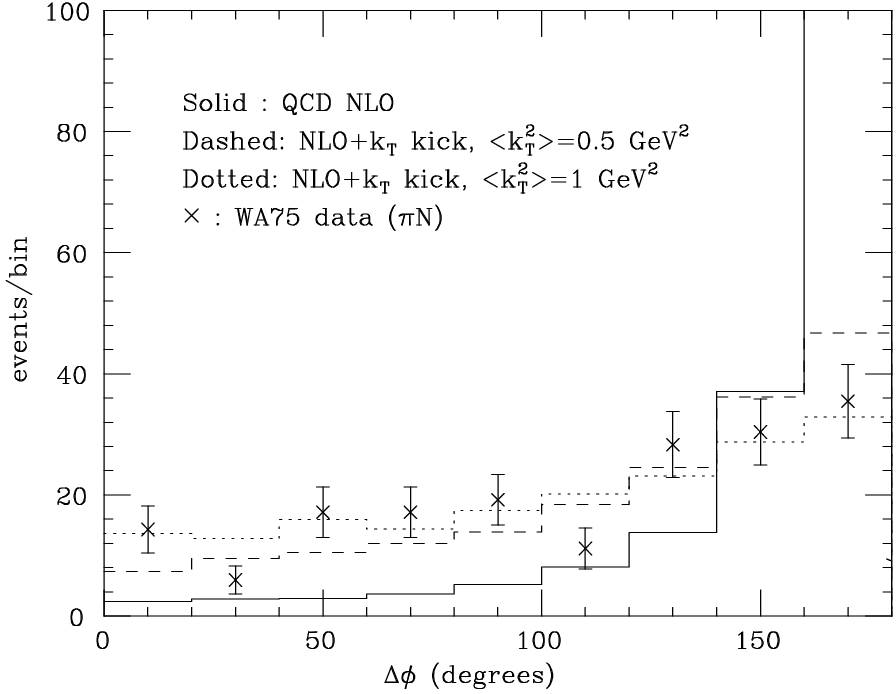,width=0.45\textwidth,clip=}
            \hspace{0.3cm}
            \epsfig{figure=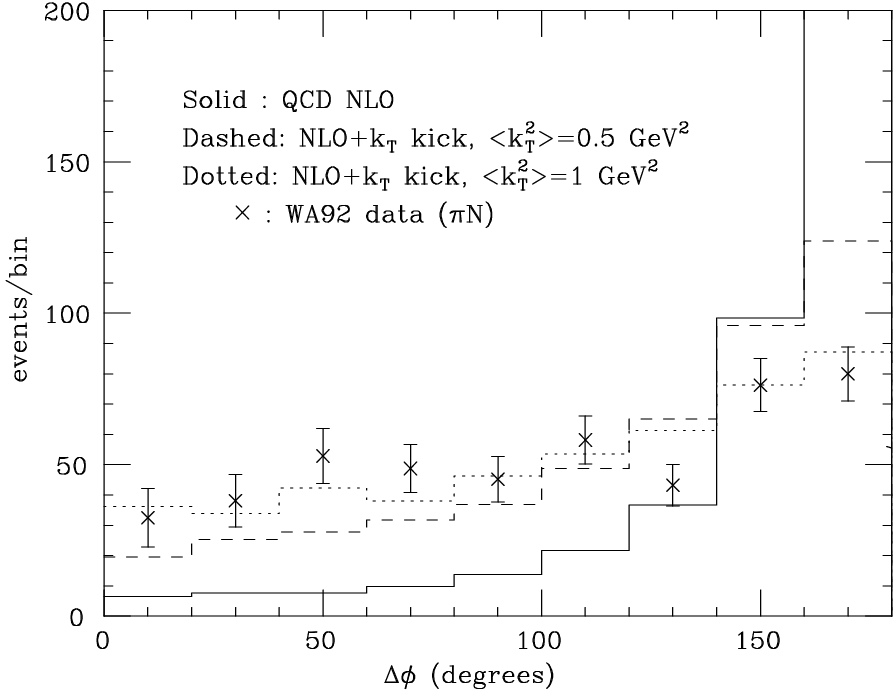,width=0.45\textwidth,clip=}}
Fig. 4. Azimuthal correlation for charm production in $\pi N$
collisions: NLO parton model and $k_T$ kick calculations versus the WA75
and WA92 data.
\end{center}
\end{figure}

The dashed and dotted histograms in Fig. 4 correspond to the NLO
parton model prediction, supplemented with the $k_T$ kick, with
$\langle k_T^2\rangle=0.5$ GeV$^2$ and $\langle k_T^2\rangle=1$ GeV$^2$,
respectively. We see that with $\langle k_T^2 \rangle= 1$ GeV$^2$ it
is possible \cite{FMNR} to describe the data.

However the large intrinsic transverse momentum significantly changes
one-particle $p_T$-distri\-butions of heavy flavour hadrons, which were
in good agreement with the data. The solid curves in Fig. 5 taken from 
\cite{FMNR} represent the NLO parton model predictions for charm quarks 
$p_T$-distributions which are in agreement with the data. The effect of 
the $k_T$ kick results in a hardening of the $p_T^2$ spectrum. On the 
other hand, by combining the $k_T$ kick with $\langle k_T^2 \rangle =1$ 
GeV$^2$ and the Peterson fragmentation \cite{Pet}, the theoretical 
predictions slightly undershoot the data (dot-dashed curves).

\begin{figure}
\begin{center}
\centerline{\epsfig{figure=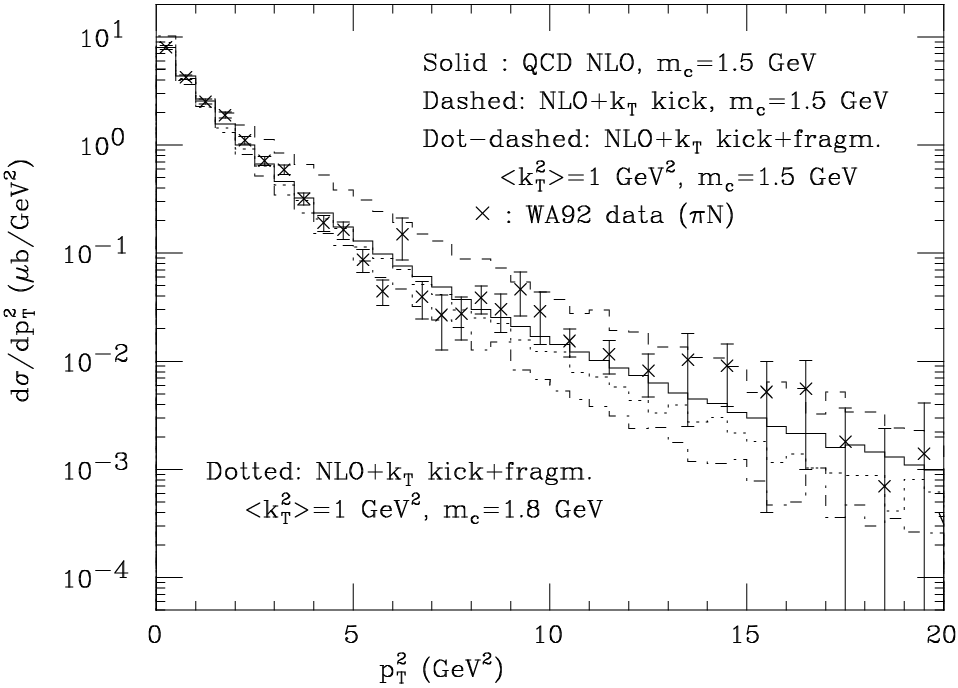,width=0.45\textwidth,clip=}
            \hspace{0.3cm}
            \epsfig{figure=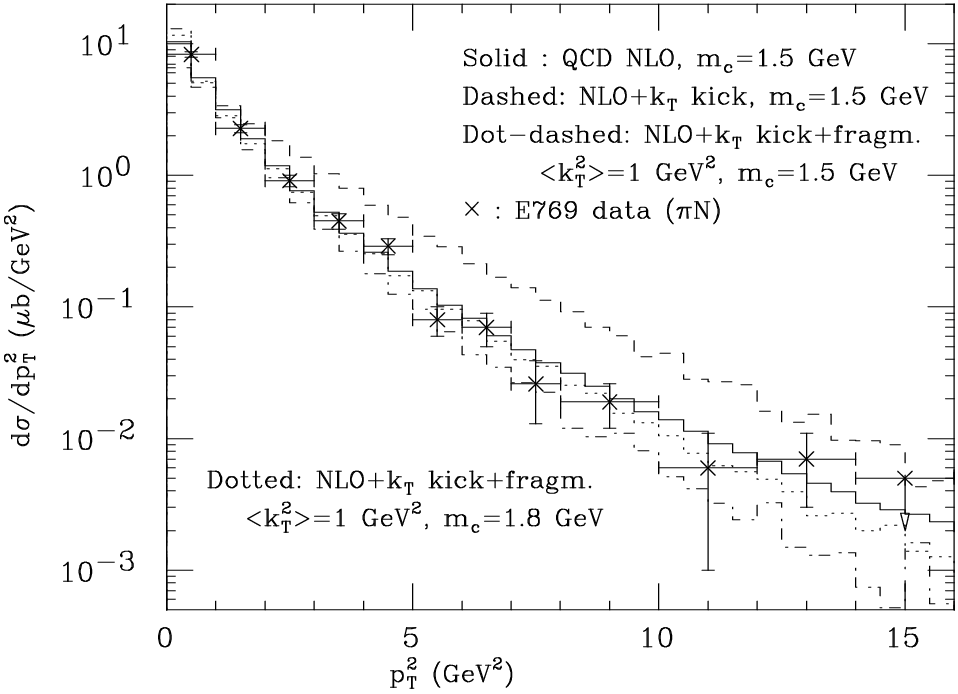,width=0.45\textwidth,clip=}}
Fig. 5. Charm $p_T^2$ distribution measured by WA92 and E769, compared
to the NLO parton model predictions, with and without the
non-perturbative effects.
\end{center}
\end{figure}

The $k_T$ kick can only very weakly change the $x_F$-distributions of
produced $c$-quarks, Fig. 3, and after accounting the fragmentation
these distributions can become too soft.

Let us consider why the conventional NLO parton model with collinear
approximation works reasonably for one-particle diustributions, and, at
the same time, it is in disagreement with the data on azimuthal
correlations.

The contribution of the processes of Fig. 1, which governs the heavy
quark production can be written\footnote{We omit for simplicity all
factors which are non-essential here.} as a convolution of initial
transverse momenta distributions, $I(q_{1T})$ and $I(q_{2T})$, with
squared modulo of perturbative QCD matrix element,
$\vert M(q_{1T}, q_{2T}, p_{1T}, p_{2T})\vert ^2$ :
\begin{equation}
\sigma_{QCD}(Q\overline{Q}) \propto
\int d^2 q_{1T} d^2 q_{2T} I(q_{1T}) I(q_{2T}) \vert M(q_{1T},
q_{2T}, p_{1T}, p_{2T}) \vert ^2 \;.
\end{equation}

Now there are two possibilities: \\
i) the essential values of initial transverse momenta are much
smaller than the transverse momenta of produced heavy quarks,
$q_{iT} \ll p_{iT}$, and \\
ii) all transverse momenta are of the same order,
$q_{iT} \sim p_{iT}$.

In LO for the first case (i) we can wait that one-dimentional $p_T$
distributions should be more broad than the distributions on the
transverse momenta of the quark pair, because in LO
$q_{1T} + q_{2T} = p_{1T} + p_{2T}$. NLO gives here a correction,
numerically not very large, see Fig. 2. In this case one can replace 
both the initial distributions $I(q_{iT})$ by $\delta$-functions,
$\delta(q_{iT})$. This reduces the expression (5) to the very
simplified one :
\begin{equation}
\sigma_{coll.}(Q\overline{Q}) \propto \vert
M(0, 0, p_{1T}, p_{2T}) \vert ^2 \;,
\end{equation}
in total agreement with Weizsaecker-Williams approximation in QED.

In the second case (ii) the distributions on the transverse momenta of
the quark pair should be the same, or even more broad than the
one-particle $p_T$-distributions, and namely this situation is realised
\cite{FMNR}. In this case we can not wait a priory that the
Weizsaecker-Williams approximation will give good results, however it
works quite reasonably in the case of one-particle distributions. In the
case of distribution on the transverse momentum of the heavy-quark pair,
$p_T^2(Q\overline{Q})$ we measure (only approximately in NLO) the
distribution over the sum of transverse momenta of incident gluons. In
this case it seems to be senseless to replace the distributions
$I(q_{1T})$ and $I(q_{2T})$ by $\delta$-functions, and to expect a
reasonable agreement with the data. The same can be said about the
azimuthal correlations.

The $k_t$ kick \cite{FMNR} effectively accounts for the transverse
momenta of incident partons. It uses the expression which can be written
symbollically as
\begin{equation}
\sigma_{kick}(Q\overline{Q}) \propto I(q_{1T}) I(q_{2T}) \otimes
\vert M(0, 0, p_{1T}, p_{2T}) \vert ^2 \;,
\end{equation}
and the main difference from the general QCD expression Eq. (5) is that 
due to absence of $q_{iT}$ in the matrix element the values of
$\langle k_T^2\rangle$ in Eq. (4) should be different for different
processes and kinematical regions. The reason is that in Eq. (5) the 
values $I(q_{iT})$ decrease at large $q^2_{iT}$ as a weak power (see 
next Sect.), i.e. comparatively slowly, and more important is the 
$q^2_{iT}$ dependence of the matrix element. In the last one the 
corrections of the order of $q^2_{iT}/\mu^2$, where $\mu^2$ is the QCD 
scale, are small enough when $q^2_{iT}/\mu^2 << 1$ and they start to 
suppress a matrix element value when $q^2_{iT}/\mu^2 \sim 1$.

\section{Heavy quark production in semihard approximation}

Let us consider another approach, when the transverse momenta of
incident gluons in the small-$x$ region appear from the diffusion of
transverse momenta in the gluon evolution\footnote{The similar approach 
based on $k_T$-factorization formulae can be found, for example, in 
Refs. \cite{CCH,CE,MW}.}. This diffusion is described by the 
function $\varphi(x,q^2)$ determined \cite{GLR} as
\begin{equation}
\label{xg}
\varphi (x,q^2) = 4\sqrt{2}\,\pi^3 \frac{d\,[xG(x,q^2)]}{d q^2} \;, 
\end{equation}
where $G(x,q^2)$ is usual gluon structure function. 

In principle the function $\varphi(x,q^2)$ which determine the 
probability to find gluon with fixed value of transverse momentum, 
$q_T$, depends on three variables, $x$, $q_T$ and gluon virtuality 
$q^2$. However at small $x$ in LLA $q_T^2 \approx - q^2$, and it leads 
to comparatively weak dependence of $\varphi(x,q^2)$ on $q_T^2$ 
(strongly different from exponential dependence in Eq. (4)) due to weak 
$q^2$-dependences of phenomenological structure functions. 

The exact expression for gluon $q_T$-distributions can be obtained, as a
solution of the nonlinear evolution equation. The calculations 
\cite{Blu} result in difference from our $\varphi(x,q^2)$ function only 
about 10-15\%.

The matrix element $M_{QQ}$ accounting for the gluon virtualities and 
polarizations is much more complicate than the parton model one. That 
is why we consider only LO contribution of the subprocess 
$gg \to Q\bar{Q}$. The differential cross section of heavy quarks 
hadroproduction has the form
$$ \frac{d\sigma_{pp}}{dy^*_1 dy^*_2 d^2 p_{1T}d^2
p_{2T}}\,=\,\frac{1}{(2\pi)^8}
\frac{1}{(s)^2}\int\,d^2 q_{1T} d^2 q_{2T} \delta (q_{1T} +
q_{2T} - p_{1T} - p_{2T}) $$
\begin{equation}
\label{spp}
\times\,\,\frac{\alpha_s(q^2_1)}{q_1^2} \frac{\alpha_s (q^2_2)}{q^2_2}
\varphi(q^2_1,y)\varphi (q^2_2, x)\vert M_{QQ}\vert^2.
\end{equation}
Here $s = 2p_a p_b\,\,$ and $y^*_{1,2}$ are the quarks' rapidities in 
the hadron-hadron c.m.s. frame. 

Eq. (9) enables to calculate straightforwardly all distributions
concerning heavy flavour one-particle, or pair production. However there 
exists a problem coming from infrared region. Gluon structure 
function in Eq. (8) is not determined at small virtualities, so 
the function $\varphi (x,q^2_2)$ is unknown at 
the small values of $q^2_2$ and $q^2_1$. To solve this problem we will
use the direct consequence of Eq. (8) \cite{Kwi}
\begin{equation}
xG(x,q^2) = xG(x,Q_0^2) + \frac{1}{4\sqrt{2}\,\pi^3}
\int_{Q_0^2}^{q^2} dq_1^2 \varphi (x,q_1^2) \;,
\end{equation}
and rewrite \cite{3} the integrals in the Eq. (9) as
the sum of four contributions. The first one is determined by the
product of two gluon distributions, $G(x,Q_0^2)$ and $G(y,Q_0^2)$,
and it is the same as the conventional LO parton model expression. Next 
three terms contain the corrections to the parton model. If the initial 
energy is not high enough, the first term dominates. In the case of very 
high energy the first term can be considered as a small corrections, and 
our results are differ from the conventional ones. In the cases when the 
collinear approximation is available, our results only slightly differ 
from the parton model.



\begin{figure}[htb]
\begin{center}
\mbox{\psfig{file=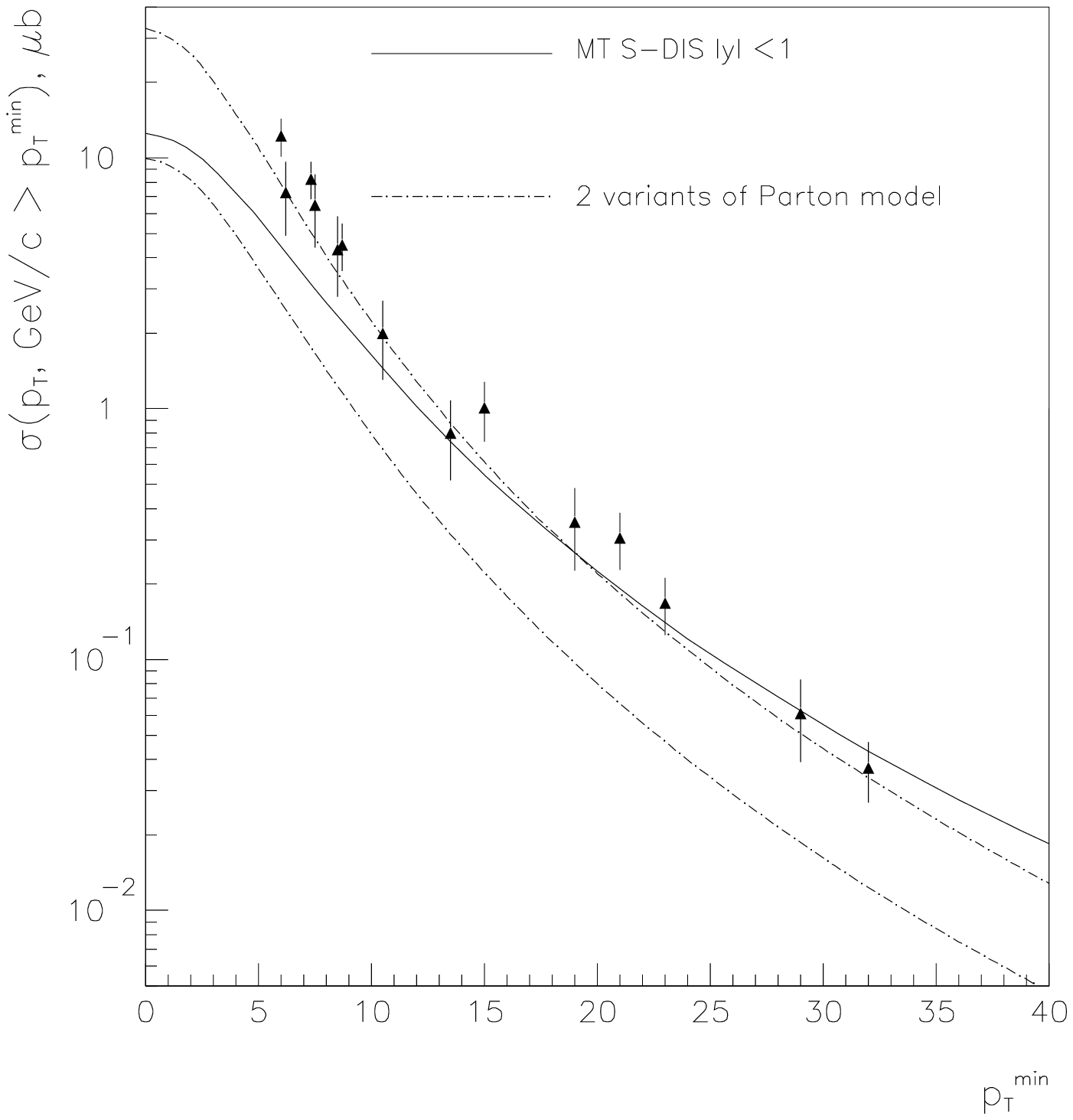,width=0.50\textwidth}} \\
Fig. 6. Cross section of beauty production in CDF.
\end{center}
\end{figure}

It is illustrated in Fig. 6 taken from \cite{3}, where we compare our 
calculations of $b$-quark $p_T$-distributions with the experimental 
results of CDF collaboration and with two variants of parton model 
calculations. The distributions over the azimuthal angle $\phi$ can be 
found in \cite{SS}.

The essential values of $q_{1T}$ and $q_{2T}$ in our calculations 
increase with increase the value of $p_T^{min}$ of detected 
$b$-quark. In the language of $k_T$ kick it means that the values of 
$\langle k_T^2\rangle$ will be also increased.

\section{Conclusion}

The experimental results on total cross sections for charm and beauty
production are in agreement with the conventional parton model 
predictions, using reasonable values of QCD scales and quark masses. The 
data on $x_F$ and $p_T$ distributions are also in reasonable agreement 
with parton model without any fragmentation functions\footnote{The $p_T$ 
distributions of charm photoproduction measured by E691 Coll. 
\cite{E691a} are more hard than the NLO parton model predictions 
\cite{FMNR,Man97}, that can be considered \cite{Man97} as an argument 
for including a non-perturbative fragmentation function. However, the 
$p_T$ slope of the calculated spectrum depends strongly on the charm 
quark mass, and the measured charm production cross section has very 
strange energy dependence \cite{E691a}.}. 

Moreover, the shapes of one-particle LO and NLO distributions
practically coinside. It means that instead of calculation the NLO
contributions, it is enough to calculate only LO contributions, and
rescale them using K-factor taken, say, from the calculated ratio of
total cross sections.

In the case of distribution over the total transverse momentum of the
produced quark pair, or azimuthal correlations, the conventional NLO 
parton model with collinear approximation can not describe the data. 
The $k_T$ kick \cite{FMNR} allows one to describe these data, however 
the problems with one-particle $p_T$-distributions appear, which can be 
solved by introducing the fragmentation function. The last way should 
produce the problems in description of $x_F$-spectra. Moreover, it 
seems that the $\langle k_T^2\rangle$ values should depend on the 
process and the kinematical regions. 

Another possibility to solve the problems of initial transverse 
momenta is to use semihard theory, accounting for the virtual nature 
of the interacting gluons, as well as their transverse motion and 
different polarizations. It results in a qualitative differences with 
the LO parton model predictions \cite{3,SS}. In Ref. \cite{RSS1} the 
values of $F_2(x,Q^2)$ were calculated using phenomenological gluon 
structure functions, and the infrared contributions to $F_2(x,Q^2)$ were 
investigated in details. The possible estimations of the shadow 
corrections in the processes of heavy flavour production can be found in 
Refs. \cite{3,LRS}.

I am grateful to M.G.Ryskin and A.G.Shuvaev for multiple discussions, 
to M.L.Mangano for very useful critical comments and to E.M.Levin who 
participated at the early stage of this activity. I thank the Organizing 
Commitee of HERA Monte Carlo Workshop for financial support. This work 
is supported by grant NATO OUTR.LG 971390.


\end{document}